\documentclass{article}
\usepackage{spconf,amsmath,graphicx}
\usepackage{xcolor}
\usepackage{multirow}
\usepackage{array}
\usepackage{makecell}
\usepackage{amsfonts}
\usepackage[percent]{overpic} 
\usepackage{booktabs}
\usepackage{colortbl}
\usepackage{hyperref}
\usepackage{enumitem}
\setlist{nosep, leftmargin=14pt}

\usepackage{mwe} 
\usepackage{enumitem}

\title{Accelerating Variational Cone-Beam CT Reconstruction with Learned Warm Starts}
\name{Héloïse Dudoignon¹, Clément Delmas¹, Laurent Facq¹, Clair Poignard², Baudouin Denis de Senneville¹}
\address{¹ University of Bordeaux, CNRS, INRIA, Bordeaux INP, IMB, UMR 5251, F-33405, Talence, France\\
² AIMOKA: Inria, AP-HP, Université Sorbonne Paris Nord, F-35042, Rennes, France\\
{\small\textit{This paper has been submitted to IEEE ISBI 2027.}}
}

\begin{document}
\ninept
\maketitle
\begin{abstract}
Variational reconstruction provides a rigorous framework for cone-beam computed tomography (CBCT) by combining an explicit acquisition model with image priors. However, iterative optimization remains computationally demanding, particularly under severe noise and sparse-view conditions. Although learning-based methods achieve high reconstruction quality, they often replace the original variational formulation with learned optimization schemes or nonlinear equilibrium models.

We propose a neural network-guided warm-start framework that accelerates variational reconstruction while preserving its original formulation. Projection data are first reconstructed using the Feldkamp–Davis–Kress (FDK) algorithm, and the resulting image is refined by an image-domain neural network. The network output is used solely to initialize the subsequent model-based optimization, leaving the forward model, data-fidelity term, regularization, and optimization scheme unchanged. A pre-regularization step further improves compatibility with the variational priors.

Experiments on pelvic CT images from the SynthRAD2025 challenge show a more than threefold reduction in reconstruction time while maintaining comparable image quality.
\end{abstract}
\section{Introduction}
\label{sec:intro}

Cone-beam computed tomography (CBCT) is widely used for image-guided interventions \cite{Rajagopal2016, Gallinato2019, Siewerdsen2011,Sutter2025_2} because it provides three-dimensional anatomical information directly in the operating room. To reduce radiation exposure, clinical protocols increasingly rely on low-dose acquisitions, sparse-view sampling, or limited-angle trajectories. These acquisition settings, however, make the reconstruction problem severely ill-posed, leading to increased noise, streak artifacts, and metal-induced artifacts.\\
\indent Model-based variational reconstruction is widely regarded as the reference approach for tackling these challenges. By combining an accurate physical acquisition model with explicit image priors, it produces reconstructions that remain consistent with the measured projection data while effectively suppressing artifacts. However, solving the associated optimization problem typically requires hundreds of iterations, limiting its applicability in time-sensitive clinical settings.\\
\indent Recent deep learning methods have substantially accelerated tomographic reconstruction. Existing approaches generally fall into three categories: supervised reconstruction networks that directly predict the reconstructed image \cite{Lu2021}, Plug-and-Play \cite{Dutta2023,Tatachak2026} or Gradient Step and Denoiser methods that replace the proximal operator with a learned denoiser, and deep unfolding methods that learn the optimization algorithm itself \cite{Moriakov2023}. While these approaches have demonstrated impressive reconstruction quality, they generally alter the original reconstruction process by replacing the explicit variational formulation with either a learned optimization scheme or a nonlinear equilibrium problem.\\
\indent The proposed approach combines deep learning with a conventional model-based reconstruction framework by using the neural network solely to initialize the iterative reconstruction. Starting from a conventional Feldkamp--Davis--Kress (FDK) reconstruction \cite{FDK}, the neural network (NN) generates an approximate solution of the reconstruction process. This estimate is then passed through a fast and lightweight pre-regularization step to ensure compatibility with the assumptions of the selected image prior. The resulting image serves as a warm-start (WS) for the subsequent variational optimization \cite{Sambharya20024}. Importantly, the learning-based components do not modify the reconstruction model itself. The final image is recovered by minimizing the original variational objective under the same physical forward model, regularization, and optimization scheme as in the corresponding classical model-based approach. The role of deep learning is therefore limited to providing a data-driven initialization that can bring the iterative solver closer to the solution, while the final reconstruction remains governed by the model-based formulation and the measured projection data. The main contributions of this work are:
\begin{itemize}
\item We propose a neural network-guided warm-start framework for variational CBCT reconstruction, in which a neural network refines a conventional FDK reconstruction to provide an effective initialization for the subsequent model-based optimization. In contrast to end-to-end learned reconstruction approaches, the proposed framework retains the original physical forward model, data-fidelity term, image prior, and optimization algorithm.
\item We introduce a pre-regularization step that promotes compatibility between the network-refined initialization and the explicit variational regularization structure, enabling the learned initialization to be integrated into the original reconstruction framework without modifying the underlying variational model.
\item We evaluate the proposed warm-start strategies on pelvic CT images from the SynthRAD2025 challenge, assessing their convergence speed, reconstruction quality, and computational cost relative to the conventional FDK initialization.
\end{itemize}
\section{Materials and Methods}
\label{sec:methods}

\subsection{Variational reconstruction model}
\label{subsec:variational}

The discrete CBCT reconstruction problem is considered, with \(\mathcal{X} = \mathbb{R}^{N}\) and \(\mathcal{Y} = \mathbb{R}^{M}\) denoting the discrete image and projection data spaces, respectively, where \(N\) and \(M\) represent the numbers of voxels and projection measurements. Let $H:\mathcal{X}\rightarrow\mathcal{Y}$ denote the CBCT forward projection \cite{Savanier2022,FriotGiroux}
operator and let $y\in\mathcal{Y}$ denote the measured projection data. The
reconstructed image $x$ is obtained by solving the minimisation problem \footnote{$||x||_p=(\sum_{i} |x_i|^p)^{\frac{1}{p}} , \quad 1 \leq p \leq \infty$}:

\begin{equation}
x^\star =
\arg\min_{x \in \mathcal{X}}
\left\{
\frac{1}{2}\|Hx-y\|_2^2 + R(x)
\right\},
\label{eq:energy_def}
\end{equation}
where the first term enforces consistency with the measured projection data
and $R(x) : \mathcal{X} \rightarrow \mathbb{R}^+$ incorporates prior information about the reconstructed image. We
use the following composite regularization term \cite{Savanier2022,Melot2012}:

\begin{equation}
R(x)
=
\frac{k}{2}\|x\|_{\mathcal{H}^1}^2
+
\rho\|Wx\|_1,
\label{eq:regularization}
\end{equation}
where $W$ denotes a transform promoting a sparse representation of the image,
and $k$ and $\rho$ control the contributions of the $\mathcal{H}^1$ and sparsity terms,
respectively. The discrete $\mathcal{H}^1$-norm is defined as:

\begin{equation}
\|x\|_{\mathcal{H}^1}^2
=
\|x\|_2^2+\|\nabla x\|_2^2, \quad x \in \mathcal{X}.
\end{equation}


\subsection{Warm start strategies}
\label{subsec:framework}

The objective of this study is to evaluate whether a learned image-domain
initialization can accelerate variational CBCT reconstruction, and whether
pre-regularizing the learned initialization further improves convergence.
Three initialization strategies are therefore compared within the same
variational reconstruction framework:

\begin{enumerate}
    \item \textbf{FDK warm start}: the analytical FDK reconstruction is directly
    used to initialize the variational optimization;
    
    \item \textbf{NN warm start}: the FDK reconstruction is processed by a
    neural network, and the resulting prediction is directly used as the
    initialization;
    
    \item \textbf{Regularized NN warm start}: the neural network prediction is
    first pre-regularized using the same priors as the variational model, and
    the resulting image is then used as the initialization.
\end{enumerate}

For all three strategies, the same variational objective, optimization
algorithm, regularization parameters, step size, and stopping criteria are
used. Thus, the initialization is the only variable modified between the
three experimental conditions.

\subsubsection{FDK warm start (FDK-WS)}
\label{subsec:fdk_warm_start}

For the analytical initialization, the low-dose projection data are
reconstructed using the FDK algorithm \cite{FDK}. Let $H^\dagger : \mathcal{Y} \rightarrow \mathcal{X}$ the corresponding analytical backprojection operator, $\mathcal{F}: \mathcal{Y} \rightarrow \mathcal{Y}$ the Ramp filter \cite{Dubois1998} and $\omega$ denote a pre-weighting factor, with $\omega y \in \mathcal{Y} $.
The FDK reconstruction is given by:

\begin{equation}
x_{\mathrm{FDK}} = H^\dagger [\mathcal{F}(\omega y)].
\label{eq:fdk}
\end{equation}

The resulting image is directly used as the initial estimate of the
variational optimization, $ x^{(0)} = x_{\mathrm{FDK}}$. No additional processing is applied to the FDK initialization.

\subsubsection{Neural network warm start (NN-WS)}
\label{subsec:nn_warm_start}

A neural network is used to improve the analytical reconstruction before
variational refinement. Let $G_\theta : \mathcal{X} \rightarrow \mathcal{X} $ denote an image-domain neural network
parameterized by $\theta$ \cite{Clerc2026}. The network takes the FDK reconstruction as input
and produces:

\begin{equation}
x_{\mathrm{NN}}
=
G_\theta(x_{\mathrm{FDK}}).
\label{eq:network_mapping}
\end{equation}

The network is trained using paired low-quality and reference image data.
For each training sample, the low-dose FDK reconstruction is used as input,
while the corresponding CT volume is used as the reference target.
For the NN warm-start condition, the network prediction is directly used as 
initial guess, $x^{(0)} = x_{\mathrm{NN}}$.

Thus, this condition evaluates the effect of using the learned image directly
as a warm start without modifying the variational reconstruction model.

\subsubsection{Regularized neural network warm start (Reg-NN-WS)}
\label{subsec:regularized_nn}

Although the neural network prediction provides an improved image-domain
estimate, it is not explicitly constrained to satisfy the regularization
structure of Eq.~\eqref{eq:regularization}. We therefore introduce a
pre-regularization step before the variational reconstruction.

Given the neural network prediction $x_{\mathrm{NN}}$, the regularized
initialization is obtained by solving:

\begin{equation}
x^{(0)}
=
\arg\min_{x \in \mathcal{X}}
\left\{
\frac{1}{2}\eta\|x-x_{\mathrm{NN}}\|_2^2
+
\frac{k}{2}\|x\|_{\mathcal{H}^1}^2
+
\rho\|Wx\|_1
\right\}.
\label{eq:pre-reg}
\end{equation}

The first term preserves the information provided by the neural network,
whereas the remaining terms impose the same smoothness and sparsity priors as
those used in the final variational reconstruction. This step therefore
adapts the learned prediction to the regularization structure of the
variational model. The first term is normalized by $\eta = \frac{|\mathcal{Y}|}{|\mathcal{X}|}. \text{max}(y)$, where $|\mathcal{Y}|$ represents the number of pixels in the sinogram 
(product of the number of projections and the number of detector cells) and $|\mathcal{X}|$ 
represents the number of pixels in an image.
This normalization ensures that the relative proportion of the two terms is similar in both 
minimization problems, compensating for the fact that the two data fidelity 
terms lie in different spaces.

Importantly, Eq.~\eqref{eq:pre-reg} does not require evaluations of the
forward projection operator $H$ or its adjoint. The pre-regularization can
therefore be performed independently of the acquisition model.

The pre-regularization problem is solved using the Forward-Backward Algorithm (FB) \cite{Dossal2024}. The objective is decomposed as:

\begin{equation}
f(x)
=
\frac{1}{2}\|x-x_{\mathrm{NN}}\|_2^2
+
\frac{k}{2}\|x\|_{\mathcal{H}^1}^2\text{, and }
\end{equation}

\begin{equation}
g(x)=\rho\|Wx\|_1.
\label{eq:fonction_g}
\end{equation}

At each iteration, the iterates are updated as follows\footnote{$\text{prox}_{\tau g} (x) = \underset{z \in \mathcal{X}}{\arg\min} \; \{\tau g(z)+ \frac{1}{2}||z-x||_2^2$\}}:

\begin{equation}
x^{(n+1)}
=
\operatorname{prox}_{\tau g}
\left(
x^{(n)}-\tau\nabla f(x^{(n)})
\right).
\end{equation}

The iterations were stopped when the relative change between two consecutive iterates fell below a prescribed tolerance $\varepsilon = 10^{-4}$.

\subsection{Variational refinement}
\label{subsec:variational_refinement}

For each initialization strategy, the image is subsequently refined by
minimizing the same variational objective given in Eq.~\eqref{eq:energy_def}.
The optimization is performed using an inertial proximal-gradient algorithm.
The objective is decomposed as:
\begin{equation}
h(x,y)
=
\frac{1}{2}\|Hx-y\|_2^2 
+
\frac{k}{2}\|x\|_{H^1}^2, \quad \text{and \eqref{eq:fonction_g}}.
\end{equation}
At iteration $n$, the iterates are updated as follows:
\begin{equation}
x^{(n+1)}
=
x^{(n)}+
\beta
\left(
\operatorname{prox}_{\tau g}(x^{(n)}-\tau \nabla_x h(x^{(n)}))-x^{(n)}
\right),
\end{equation}
where $\tau$ is the gradient step size and $\beta$ is an inertial parameter.

\subsection{Experimental setup}
\label{subsec:experimental_setup}

\subsubsection{Dataset and CBCT simulation} 
\label{subsubsec:lowdose_simulation}

Experiments were conducted using the SynthRAD2025 challenge dataset~\cite{SynthRad2023}, comprising 180 three-dimensional CT volumes. The dataset was divided into 170 training volumes and 10 held-out test volumes. The test volumes were not used during network training and were exclusively reserved for evaluation. All images were resampled to a standardized isotropic voxel size of $2.5\!\times\!2.5\!\times\!2.5$ mm.

Low-dose CBCT data were simulated from the CT volumes using a standard cone-beam acquisition geometry with 60 angular projections and a $300 \times 300$ detector. Projection data were generated from the CT volumes, and Poisson noise was introduced in the projection domain \cite{Tu2006} to simulate photon-counting noise, assuming an incident photon count of $10^8$ photons. The resulting noisy projections were reconstructed using FDK reconstruction~\cite{FDK}. The resulting FDK images were used as analytical baseline reconstructions and as inputs to the proposed neural network.

\subsubsection{Neural Network training}
\label{subsubsec:network_training}

A nnUNet v2 network $\footnote{
\url{https://github.com/phyrise/nnUNet_translation}}$, following the implementation described by Longuefosse \textit{et al.}~\cite{Longuefosse2024_adapt_nnUnet}~\cite{Longuefosse2025}, was trained to generate the learned initialization. The network was trained on the 170 training volumes using
mean squared error (MSE) loss, without intensity normalization, for 200
epochs. Low-dose FDK reconstructions were used as inputs, while the
corresponding CT volumes were used as reference images.

At inference time, the trained network was applied exclusively to the
10 held-out test volumes. The resulting predictions were used either
directly as the NN warm start or as input to the pre-regularization step for
the regularized NN warm start.

\subsubsection{Variational reconstruction parameters}
\label{subsubsec:reconstruction_parameters}

The performance of the proposed approach was investigated for two values of the sparsity parameter in Eq. \eqref{eq:regularization}, namely, $\rho \in \{10, 50\}$. The $H^1$ regularization parameter was fixed at $k=10^{-3}$ to ensure that its contribution to the overall regularization remained minor compared with that of the sparsity term. The optimization step size and inertial parameter were empirically set to $\tau=2\times10^{-5}$ and $\beta=1$, respectively. For all three initialization strategies, the same variational objective, optimization algorithm, parameter values, and stopping criteria were employed.

\subsubsection{Reference solutions and evaluation}
\label{subsec:evaluation}

For each CT volume, a reference solution $x^{*}$ was obtained by solving the corresponding variational problem using the same inertial proximal-gradient algorithm as for the evaluated reconstruction methods. Specifically, the iterations were terminated when the relative change between two consecutive iterates fell below a stringent prescribed tolerance $\varepsilon = 10^{-6}$.

Reconstruction quality of approximated solutions $\hat{x}$ was assessed using the Structural Similarity Index Measure (SSIM), which captures structural and statistical similarities between $\hat{x}$ and $x^{*}$. SSIM was evaluated at each optimization iteration to characterize the evolution of image quality throughout the iterative refinement process. Since the proposed approach aims to accelerate convergence through improved initialization, the number of optimization iterations, reconstruction time, and variational energy were also recorded. To quantify convergence speed in terms of reconstruction quality, we defined $N_\alpha$ as the first iteration at which the reconstruction reached a prescribed SSIM threshold $\alpha$:
\begin{equation}
N_\alpha =
\min \left\{
n : \mathrm{SSIM}(x^{(n)},x^{*}) \geq \alpha
\right\}.
\label{eq:nalpha}
\end{equation}

Similarly, $T_\alpha$ was defined as the minimum elapsed time required to reach the same SSIM threshold, including the computational cost of generating the initial estimate (warm start). Specifically, this overhead comprised the FDK reconstruction time for FDK-WS, neural network inference time for NN-WS, and both neural network inference and regularization of the network prediction for Reg-NN-WS. Thus, $T_\alpha$ accounts for the total computational time required to reach the prescribed reconstruction quality, including the method-specific initialization overhead. 

The evaluation was performed using SSIM thresholds of $\alpha \in \{0.95, 0.99\}$. Pairwise comparisons of the computation times required to reach each SSIM threshold were performed among the FDK-WS, NN-WS, and Reg-NN-WS strategies using paired Student’s \(t\)-tests, with statistical significance set at $p < 0.05$.

\subsubsection{Hardware and implementation}

Time benchmarks were performed on a workstation running Ubuntu 26.04 LTS and equipped with an Intel\textsuperscript{\textregistered} Xeon\textsuperscript{\textregistered} Gold Sliver 4410Y CPU. All implementations were developed in Python 3.11. Spatial gradients were computed using finite differences. CBCT projection generation and FDK reconstruction were performed using the Reconstruction Toolkit (RTK) \cite{Rit2024}. Code available after acceptance\footnote{https://github.com/xxx}.

\section{Results}
\label{sec:pagestyle}

\begin{figure}[h!]
    \centering
	\begin{minipage}[b]{0.49\linewidth}
        \centering
            \includegraphics[height=3.6cm]{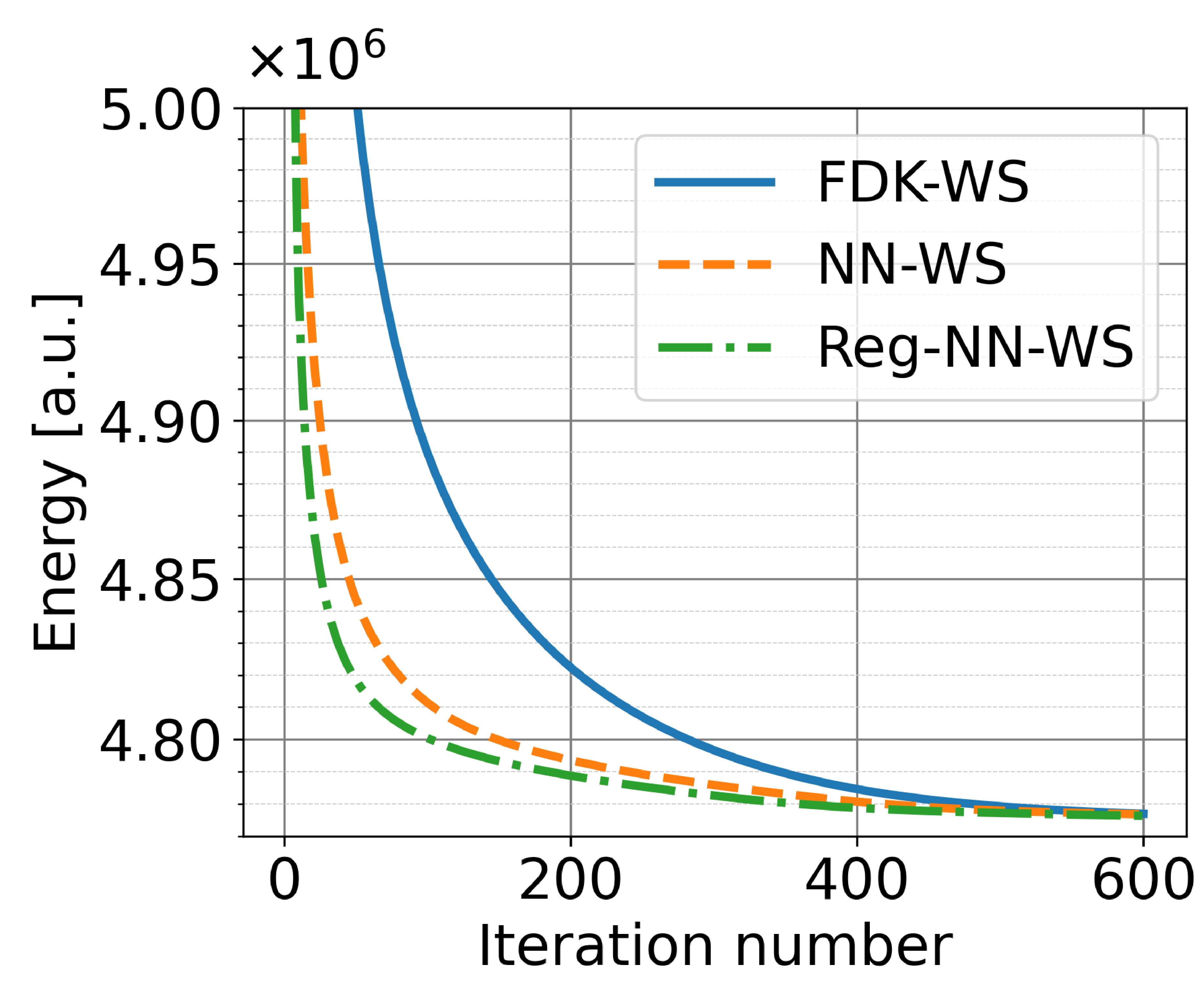}
        \centerline{(a)}
    \end{minipage}
    \hfill
    \begin{minipage}[b]{0.49\linewidth}
        \centering
            \includegraphics[height=3.4cm]{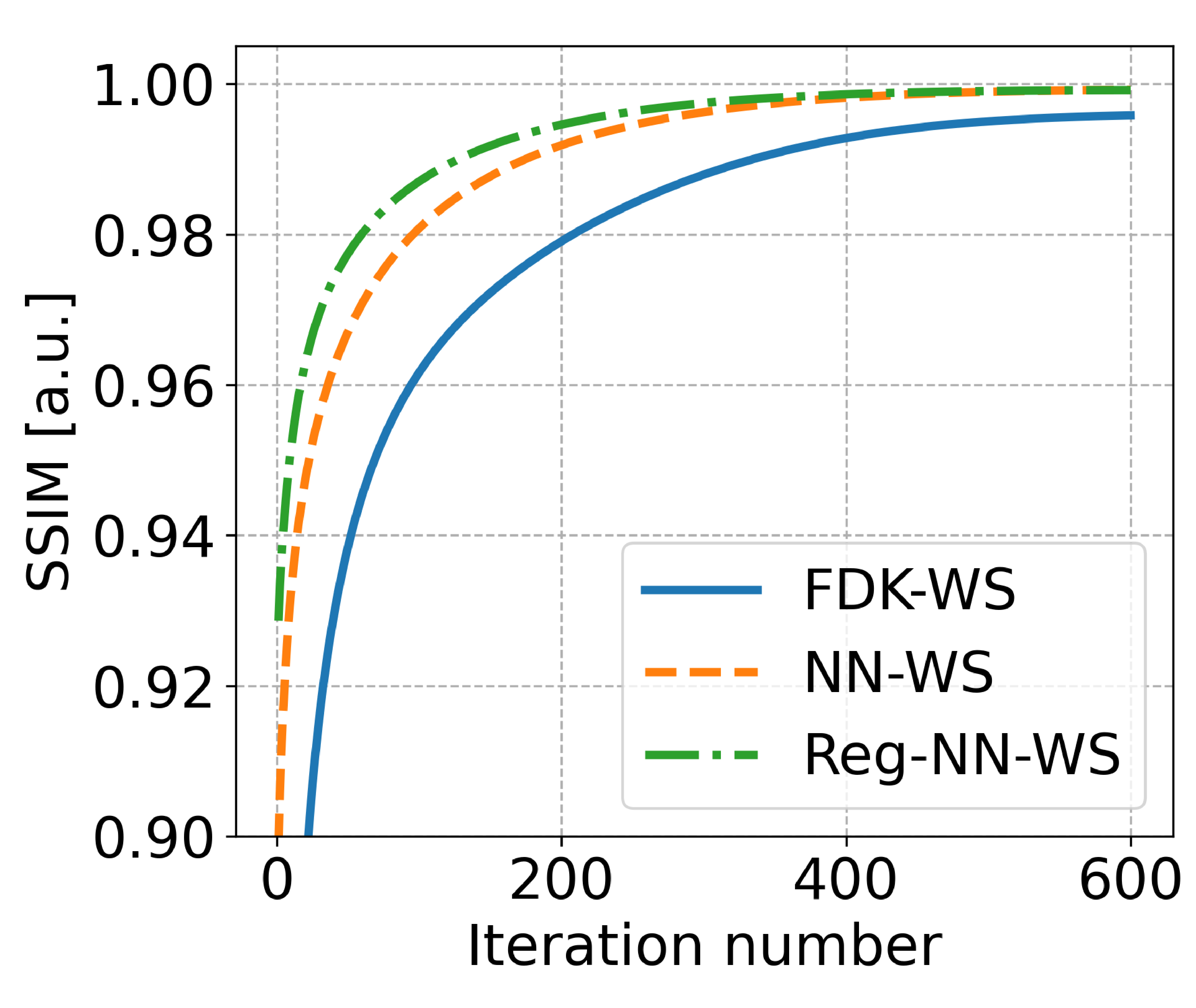}
        \centerline{(b)}
    \end{minipage}
\caption{Typical evolution of the variational energy (a) and SSIM (b) as a function of the number of iterations during variational refinement, for the three initialization strategies considered, with $\rho = 50$.}
\label{fig:res}
\end{figure}

Figure~\ref{fig:res} shows that the variational energy consistently decreases toward zero over the course of the variational refinement process for the three initialization strategies considered. Similarly, the SSIM between the reference and reconstructed solutions increases toward 1. The convergence rate is improved when using the NN-WS strategy compared with the FDK-WS approach, and is further enhanced by the Reg-NN-WS strategy, particularly during the first iterations of the refinement process. This behavior is confirmed in Table~\ref{fig:tab} for both configurations of the regularization parameter $\rho$ considered in the general formulation of the problem defined in Eq.~\eqref{eq:regularization}.

Each initialization strategy introduces a fixed computational overhead associated with the warm-start generation. The computation time required to generate the warm start was 2.7~s for the FDK-WS strategy. This initialization cost was replaced by a 1~s inference time of the nnUNet model for the NN-WS strategy. For the Reg-NN-WS approach, an additional 6.3~s was required for the pre-regularization step applied to the neural network prediction. Considering an average computation time of 1.98~s per iteration during the variational refinement process, the reduction in the number of required iterations directly translates into a substantial decrease in the total reconstruction time, as reported in Table~\ref{fig:tab}. Pairwise comparisons reveal statistically significant differences in computation time among the three initialization strategies (FDK-WS, NN-WS, and Reg-NN-WS) across the 10 held-out test volumes.

\begin{figure}[h!]
    \centering
	\begin{minipage}[b]{0.32\linewidth}
        \centering
        {\footnotesize Reference solution \\ ($x^{*}$)}\\
            \includegraphics[height=2.5cm]{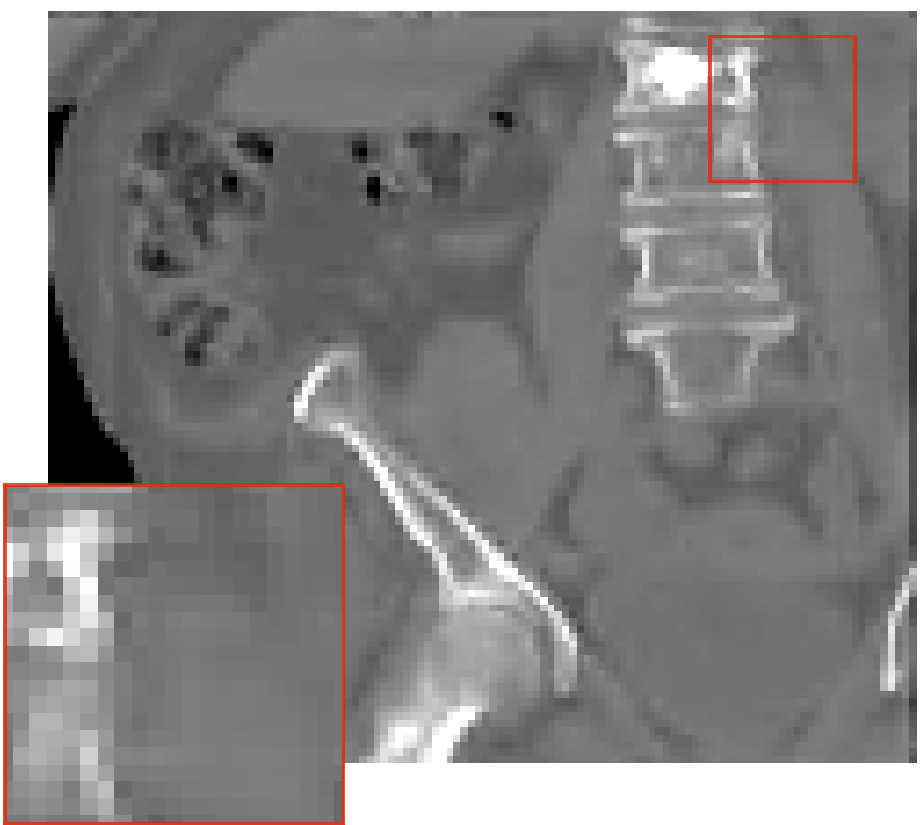}
        \centerline{(a)}
    \end{minipage}
    \hfill
    \begin{minipage}[b]{0.32\linewidth}
        \centering
        {\footnotesize FDK reconstruction \\ ($x_{FDK}$)}\\
            \includegraphics[height=2.5cm]{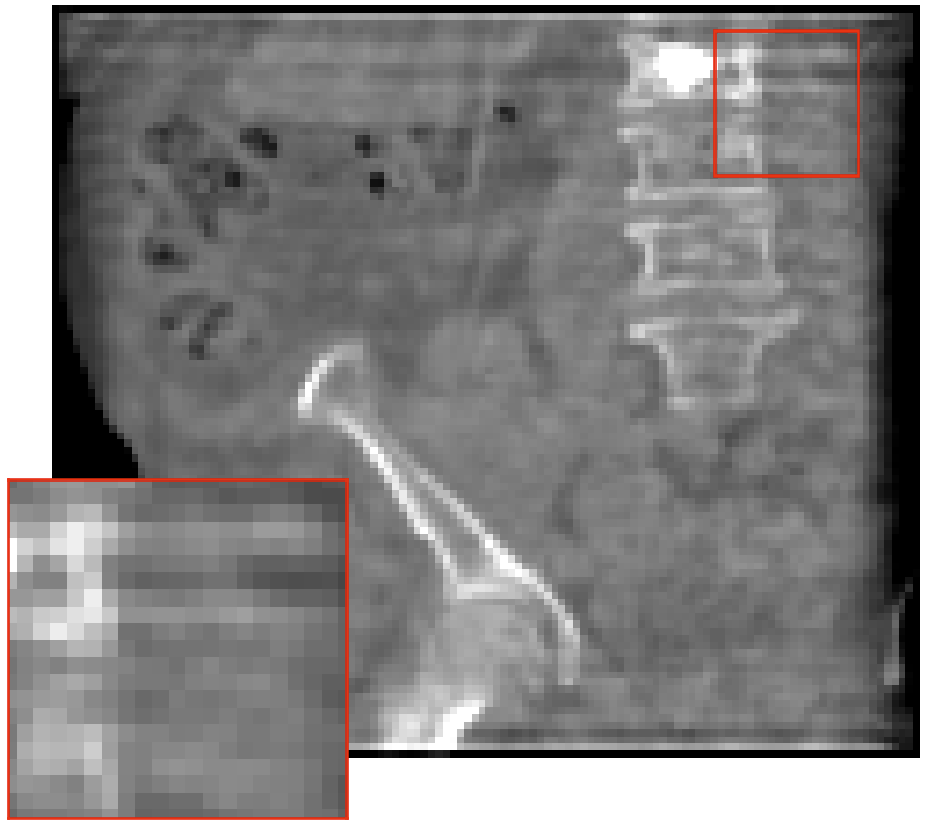}
        \centerline{(b)}
    \end{minipage}
    \hfill    
    \begin{minipage}[b]{0.325\linewidth}
        \centering
        {\footnotesize Approximate sol. ($\hat{x}$)}\\{\scriptsize [it=$N_{0.99}$/Reg-NN-WS]}\\
            \includegraphics[height=2.5cm]{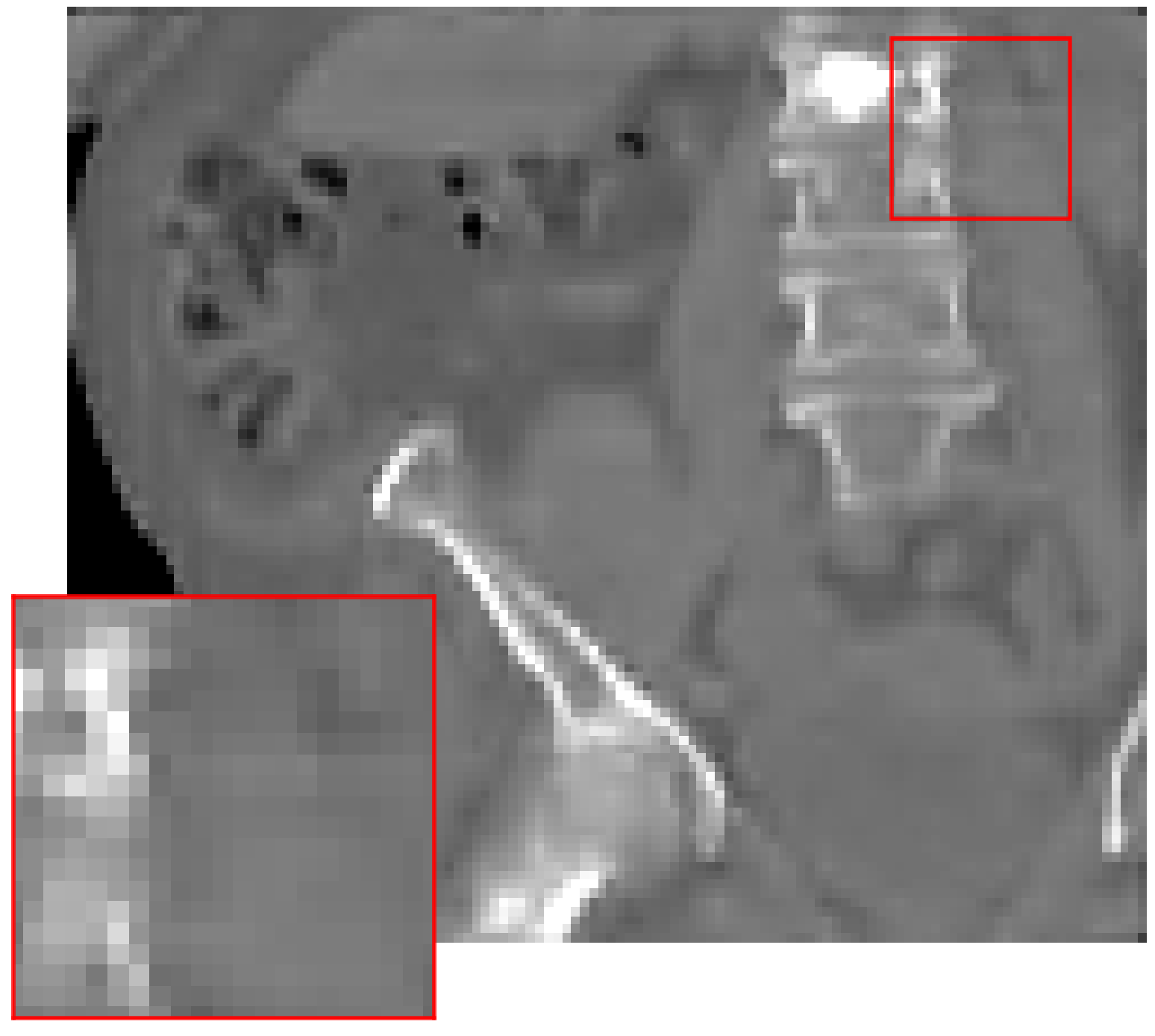}
        \centerline{(c)}
    \end{minipage}
    \caption{Typical example of a 2D pelvic image in the coronal plane: (a) reference solution; (b) conventional FDK reconstruction; and (c) approximate solution obtained using the regularized NN warm-start strategy at the iteration at which the reconstruction reached a prescribed SSIM threshold of 0.99, with $\rho=10$.}
  \label{fig:images}
\end{figure}

Figure~\ref{fig:images} illustrates a representative CBCT reconstruction obtained with the Reg-NN-WS strategy at the iteration corresponding to the achievement of the predefined SSIM threshold of 0.99. For a comparable reconstruction quality, the proposed framework enables a reduction of the overall reconstruction time by more than a factor of three using Reg-NN-WS compared with FDK-WS, as summarized in Table~\ref{fig:tab}.

\begin{table}[htb]
\centering
\scriptsize
\setlength{\tabcolsep}{3pt}
\renewcommand{\arraystretch}{1.2}
\begin{tabular}{c!{\vrule}ccc!{\vrule}ccc}
  & \multicolumn{3}{c!{\vrule}}{Weak regularization ($\rho = 10$)} 
  & \multicolumn{3}{c}{\makecell{Strong regularization ($\rho = 50$)}} \\
\cline{1-7}
WS strategy  &  FDK-WS & NN-WS & Reg-NN-WS & FDK & NN-WS & Reg-NN-WS\\
\hline
\makecell[l]{$N_{0.95} [\#]$}
& \cellcolor{orange!40}153
& \cellcolor{yellow!40}10
& \cellcolor{green!40}3
& \cellcolor{orange!40}62
& \cellcolor{yellow!40}11
& \cellcolor{green!40}2\\
\makecell[l]{$N_{0.99} [\#]$}
& \cellcolor{orange!40}829
& \cellcolor{yellow!40}167
& \cellcolor{green!40}148
& \cellcolor{orange!40}285
& \cellcolor{yellow!40}108
& \cellcolor{green!40}78\\
\hline
\makecell[l]{$T_{0.95} $ [s]}
& \cellcolor{orange!40}305
& \cellcolor{yellow!40}20
& \cellcolor{green!40}13
& \cellcolor{orange!40}125
& \cellcolor{yellow!40}23
& \cellcolor{green!40}10\\
\makecell[l]{$T_{0.99}$ [s]}
& \cellcolor{orange!40}1643
& \cellcolor{yellow!40}331
& \cellcolor{green!40}300
& \cellcolor{orange!40}566
& \cellcolor{yellow!40}214
& \cellcolor{green!40}163\\
\end{tabular}
\caption{Summary of the number of iterations and computational time required to reach the prescribed SSIM thresholds of 0.95 and 0.99, averaged across the 10 held-out test volumes, for the three initialization strategies and the two values of the sparsity parameter $\rho \in \{10, 50\}$ in Eq.~\eqref{eq:regularization}. The cells are color-coded according to their relative values within each group corresponding to a given $\rho$: green, yellow, and orange indicate the lowest, intermediate, and highest values, respectively.}
\label{fig:tab}
\end{table}

\section{Discussion and Conclusion}
\label{sec:discussion_conclusion}

Fully learned reconstruction methods offer substantial computational and methodological advantages, but their outputs are ultimately determined by the learned mapping and the distribution represented in the training data. Consequently, the reconstruction is not explicitly constrained by the physical inverse problem beyond what is implicitly encoded during training. Several recent approaches, including deep equilibrium formulations, have sought to combine the flexibility of learned models with the structure of variational or model-based reconstruction. Nevertheless, the computational and memory requirements of such methods remain important challenges, particularly for large-scale three-dimensional reconstruction \cite{Clerc2026}.

These limitations are particularly relevant in clinical applications, where training and test distributions may differ in terms of patient populations, acquisition protocols, and pathologies. This is especially challenging in interventional CBCT, where patient-specific anatomy and rare pathologies may be poorly represented in training datasets. In contrast, explicitly preserving the underlying inverse problem provides greater flexibility to adapt the reconstruction to specific clinical applications while maintaining the physical acquisition model.

In this work, we therefore investigated a complementary use of deep learning: rather than replacing the variational reconstruction, the neural network is used to provide an informed initialization of the subsequent optimization. We proposed a neural-network-guided warm-start strategy for variational CBCT reconstruction, in which learning is leveraged to accelerate the optimization while preserving the original physical and mathematical formulation. The proposed approach thus combines the computational efficiency of learned image priors with the explicit physical modeling and optimization framework of variational reconstruction.

In the present study, we demonstrate that a simple end-to-end modification of the nnU-Net architecture can provide an effective initialization for the variational reconstruction~\cite{Longuefosse2024_adapt_nnUnet}. The resulting computational benefit, however, inherently depends on the proximity of the learned initialization to the solution of the underlying variational problem. The pre-regularization step should be regarded as a heuristic initialization rather than an exact regularization of the original optimization problem. The data-fidelity term in Eq. \eqref{eq:regularization} is formulated in the image domain, unlike the sinogram-domain formulation of Eq. \eqref{eq:energy_def}, making the two formulations non-equivalent. Normalization was therefore applied to account for their different units. Despite this difference, the pre-regularization provides a computationally inexpensive and effective initialization for the subsequent variational optimization.

Overall, these results support the use of learned models as optimization accelerators rather than as substitutes for physics-based reconstruction, consistent with previous investigations in other numerical modeling contexts \cite{Desier2026}. Future work will focus on developing more solution-aware strategies for generating warm starts,  potentially leveraging physics-informed neural networks (PINNs) \cite{Lawal2022},to further improve convergence and reduce computational cost. Further studies will also assess the benefits of the proposed approach for task-specific applications, particularly segmentation and image registration, while addressing metal artifacts and evaluating its scalability to large-scale three-dimensional reconstruction \cite{Sutter2025,Haneda2025}.

\section{Compliance WITH ETHICAL STANDARDS}

The data analyzed in this study were obtained from the publicly available SynthRAD2025 challenge dataset. The use of these data follows the terms and conditions defined by the challenge organizers for scientific research purposes. No additional patient data were collected, and this study involved only the analysis of anonymized datasets provided through the challenge. Since the authors did not have access to identifiable patient information and no interaction with human participants was conducted, institutional review board approval and informed consent were not applicable to this study.

\section{ACKNOWLEDGMENTS}
The authors would like to thank Marion Savanier for her valuable support and fruitful discussions during this work.

This research has been conducted with the partial financial support of the France 2030 BPI project Cone Beam AI and the Institut National du Cancer (INCa) (PLBIO n°2023-156). B. Denis de Senneville was also partly granted by the ANR projects OPLA and ROCK-AI. C. Poignard was also partly granted by the ANR projects IMITATE and DEVIN.

During the preparation of this manuscript, the authors used Le Chat by Mistral AI, Claude by Anthropic and ChatGPT by OpenAI to refine the clarity and language of the text, and to support the development. After using these tools, the authors carefully reviewed and edited the content and take full responsibility for the final version.

\bibliographystyle{IEEEbib}
\bibliography{2026_Heloise_ISBI_low_dose_reco_v8}

\end{document}